\documentclass[twocolumn,showpacs,aps,prd]{revtex4-1}

\usepackage{graphicx}
\usepackage{dcolumn}
\usepackage{amsmath}
\usepackage{epsfig}
\usepackage{subfigure}
\usepackage{epstopdf}
\usepackage{multirow}
\usepackage{mathrsfs}
\usepackage[dvipdfmx, bookmarksnumbered, pdfstartview=FitH,colorlinks,urlcolor=blue, citecolor=blue,linkcolor=blue,] {hyperref}

\usepackage{lineno}
\usepackage{enumerate}
\usepackage{float}
\usepackage{color}

\begin{document}

\title{\bf \boldmath
 Search for baryon and lepton number violating decays of $\Xi^{0}$ hyperons
}

\author{
\begin{small}
\begin{center}
M.~Ablikim$^{1}$, M.~N.~Achasov$^{13,b}$, P.~Adlarson$^{75}$, R.~Aliberti$^{36}$, A.~Amoroso$^{74A,74C}$, M.~R.~An$^{40}$, Q.~An$^{71,58}$, Y.~Bai$^{57}$, O.~Bakina$^{37}$, I.~Balossino$^{30A}$
% ...
\\
\vspace{0.2cm}
(BESIII Collaboration)\\
\vspace{0.2cm} {\it
$^{1}$ Institute of High Energy Physics, Beijing 100049, People's Republic of China\\
% ...
$^{13}$ G.I. Budker Institute of Nuclear Physics SB RAS (BINP), Novosibirsk 630090, Russia\\
% ...
$^{30}$ INFN Sezione di Ferrara, (A)INFN Sezione di Ferrara, I-44122, Ferrara, Italy; (B)University of Ferrara,  I-44122, Ferrara, Italy\\
% ...
$^{36}$ Johannes Gutenberg University of Mainz, Johann-Joachim-Becher-Weg 45, D-55099 Mainz, Germany\\
$^{37}$ Joint Institute for Nuclear Research, 141980 Dubna, Moscow region, Russia\\
% ...
$^{40}$ Liaoning Normal University, Dalian 116029, People's Republic of China\\
% ...
$^{57}$ Southeast University, Nanjing 211100, People's Republic of China\\
$^{58}$ State Key Laboratory of Particle Detection and Electronics, Beijing 100049, Hefei 230026, People's Republic of China\\
% ...
$^{71}$ University of Science and Technology of China, Hefei 230026, People's Republic of China\\
% ...
$^{74}$ University of Turin and INFN, (A)University of Turin, I-10125, Turin, Italy; (B)University of Eastern Piedmont, I-15121, Alessandria, Italy; (C)INFN, I-10125, Turin, Italy\\
$^{75}$ Uppsala University, Box 516, SE-75120 Uppsala, Sweden\\
% ...
$^{b}$ Also at the Novosibirsk State University, Novosibirsk, 630090, Russia\\
% ...
}\end{center}
\vspace{0.4cm}
\end{small}
}

\begin{abstract}
Using $(1.0087\pm0.0044)\times10^{10}$ $J/\psi$ events collected by the BESIII detector at the BEPCII collider, we report the first search for the baryon and lepton number violating decays $\Xi^{0} \rightarrow K^{-} e^{+}$ with $\Delta(B-L)=0$ and $\Xi^{0} \rightarrow K^{+} e^{-}$ with $|\Delta(B-L)|=2$, where $B$ ($L$) is the baryon (lepton) number. While no signal is observed, the upper limits on the branching fractions of these two decays are set to $\mathcal B(\Xi^{0} \rightarrow K^{-} e^{+})<3.6\times10^{-6}$ and $\mathcal B(\Xi^{0} \rightarrow K^{+} e^{-})<1.9\times10^{-6}$ at the 90\% confidence level, respectively. These results offer a direct probe of baryon number violating interactions involving a strange quark.
\end{abstract}

\maketitle

\section{\boldmath INTRODUCTION}
\label{introduction}
The matter-antimatter asymmetry observed in the universe is one of the major frontier issues to be solved in particle physics, nuclear physics, and astrophysics. Astronomical observations suggest that the number of baryons in the universe far exceeds the number of antibaryons~\cite{1997AdamsRMP69}. To explain this asymmetry, Sakharov proposed the conservation of a baryon-muon charge which would allow for violation of baryon number conservation in the early universe~\cite{1967SakharovPZETF5}.

Many models predict baryon number violating (BNV) decays within and beyond the Standard Model (SM). One class of them with leading-order BNV effects~\cite{1979WeinbergPRL43} uses dimension-six operators which arise  with the conservation of $B-L$ symmetry when the SM is embedded in grand unified theories (GUTs) such as SU(5)~\cite{1974GeorgiPRL32}. The predicted BNV decays obey the selection rule $\Delta(B-L)=0$, where $\Delta(B-L)$ denotes the change in the difference between baryon and lepton numbers. The SU(5) GUT proposes the existence of two new gauge bosons, the $X$ with charge $\frac{4}{3}e$ and the $Y$ with charge $\frac{1}{3}e$, and allows $q\rightarrow X\ell$ vertices, where $q$ is a quark and $\ell$ is a lepton. Since the minimal SU(5) model is essentially excluded by the proton decay experiment~\cite{2009SKPRL102}, we require alternative models that allow for BNV decays but do not conflict with the present data, as {\it e.g.} the ``flipped SU(5)" model~\cite{1989IPLB231}. Another class of SM extensions with next-to-leading BNV effects uses dimension-seven operators~\cite{1980WeinbergPRD22}, arising from the spontaneous breaking of $B-L$ symmetry when the SM is embedded in GUTs such as SO(10)~\cite{2012BabuPRL109}, which can be mediated by an elementary scalar field $\phi$~\cite{1980WeinbergPRD22}.

Experimentally, potential BNV processes have been searched for in $D$ decays~\cite{2020DPRD101}, $J/\psi$ decays~\cite{2019JpsiPRD99},  $\tau$ decays~\cite{2020tauPRD102}, $B$ decays~\cite{2011BPRD83}, and top-quark decays~\cite{2014TopPLB731}. So far, no signal has been observed, and the upper limits on their branching fractions were set to be $10^{-3}\sim10^{-8}$ at the 90\% confidence level. BNV decays of the proton as the lightest baryon have been searched for for several decades with null results, and the strictest constraint on its partial lifetime was set to $\tau/\mathcal{B}(p\rightarrow \mu^{+}\pi^{0})>1.6\times10^{34}$ years at the 90\% confidence level~\cite{2020SKPRD102}. The proton stability has been used to stringently constrain BNV decays involving higher-generation quarks ($i.e.$, $c$, $b$ and $t$)~\cite{2005HouPRD72}. However, since multiple amplitudes can contribute to a given decay, these can have constructive or destructive interference according to their relative phases, a fact that the theoretical calculations that constrain the BNV decays do not consider due to the ample parameter space~\cite{2015LambdaPRD92}. This possible interference allows for observations in decay modes involving $u$ or $d$ quarks, coupled to other quark flavors, even though the BNV interactions are not occurring in that particular decay mode. For the BNV decays coupled to the strange quark in the initial states, only the CLAS experiment~\cite{2015LambdaPRD92} has searched for $\Lambda$ hyperon decays, imposing an upper limit on the branching fraction in the range of $10^{-5}\sim10^{-7}$ at the 90\% confidence level. Other BNV decays involving strange quarks have yet to be investigated. So far, the decays of $\Xi^{0}$ hyperons such as $\Xi^{0} \rightarrow K^{\pm} e^{\mp}$ with a lepton number violation as a consequence of BNV have never been searched for. In order to directly probe the BNV processes involving strange quarks in the initial states, these decays were proposed to be studied with the world's largest $J/\psi$ data sample at BESIII~\cite{2017LiFP12}.

\begin{figure}[tb]
\centering
\begin{minipage}{4.2cm}
\includegraphics[width=1\textwidth]{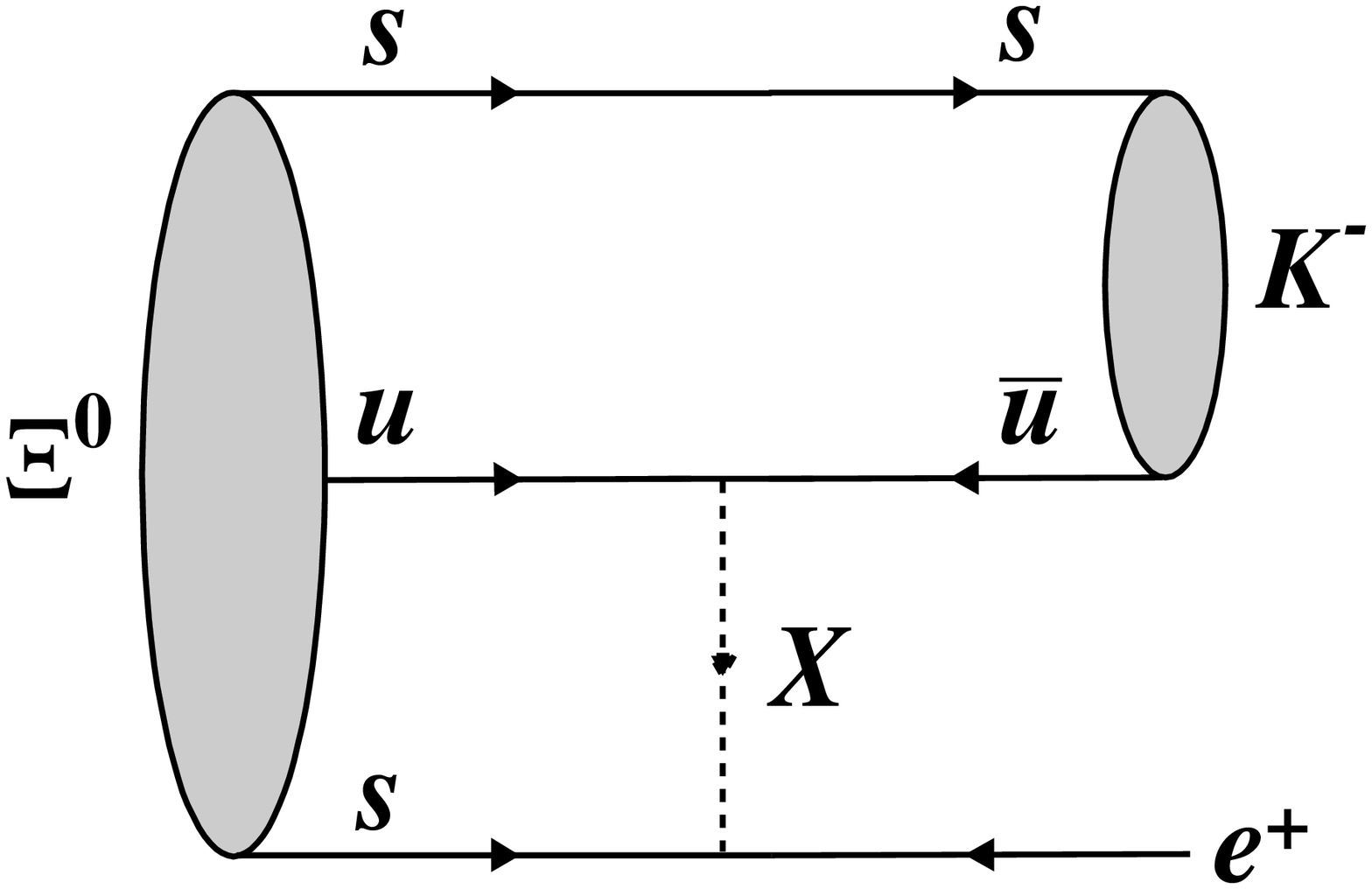}
\centerline{(a)}
\end{minipage}
\begin{minipage}{4.2cm}
\includegraphics[width=1\textwidth]{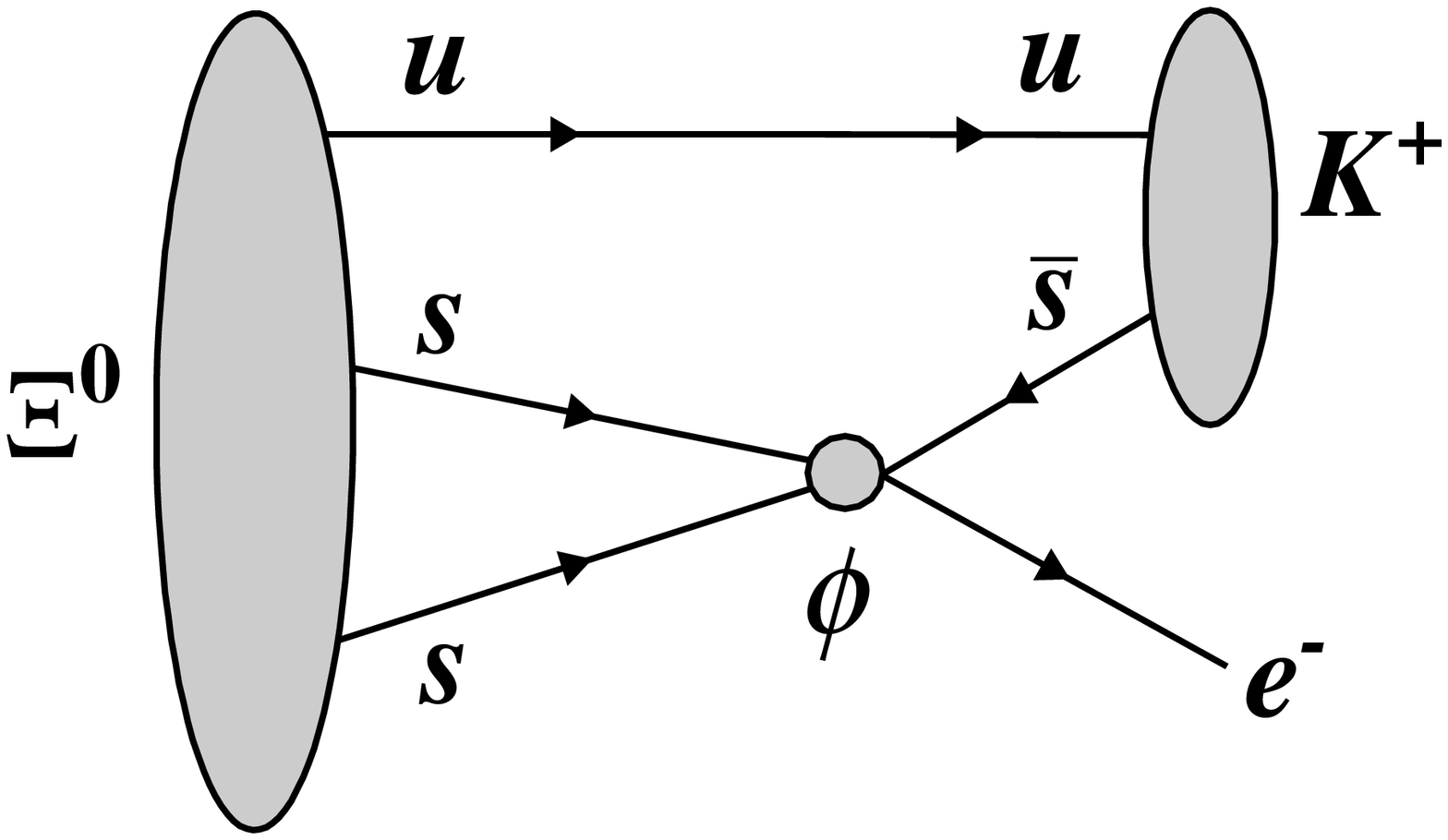}
\centerline{(b)}
\end{minipage}
\caption{Feynman diagrams for the BNV decays of (a) $\Xi^{0} \rightarrow K^{-} e^{+}$ with $\Delta(B-L)=0$ mediated by a gauge boson $X$ and (b) $\Xi^{0} \rightarrow K^{+} e^{-}$ with $|\Delta(B-L)|=2$ mediated by an elementary scalar field $\phi$.}
\label{fig:fey}
\end{figure}

In this paper, we present the first search for the baryon and lepton number violating decays of $\Xi^{0} \rightarrow K^{-} e^{+}$ with $\Delta(B-L)=0$ and $\Xi^{0} \rightarrow K^{+} e^{-}$ with $|\Delta(B-L)|=2$, as shown in the Feynman diagrams in Fig.~\ref{fig:fey}, by analyzing $(1.0087\pm0.0044)\times10^{10}$ $J/\psi$ events~\cite{2022YangCPC46} collected at a center-of-mass energy $\sqrt{s} = 3.097$ ${\rm GeV}$ with the BESIII detector. Charge-conjugate channels are always implied throughout this paper.

\section{\boldmath{DETECTOR AND MONTE CARLO SIMULATION}}
\label{detector}

The BESIII detector~\cite{2009AblikimNIMA614} records symmetric $e^+e^-$ collisions provided by the BEPCII storage ring~\cite{2016YuIPACTUYA01} in the center-of-mass energy range from 2.0 to 4.95~${\rm GeV}$, with a peak luminosity of $1 \times 10^{33}\;\text{cm}^{-2}\text{s}^{-1}$
achieved at $\sqrt{s} = 3.77\;\text{GeV}$. BESIII has collected large data samples in this energy region~\cite{2019AblikimCPC44}. The cylindrical core of the BESIII detector covers 93\% of the full solid angle and consists of a helium-based multilayer drift chamber~(MDC), a plastic scintillator time-of-flight system~(TOF), and a CsI(Tl) electromagnetic calorimeter~(EMC), which are all enclosed in a superconducting solenoidal magnet providing a 1.0 T (0.9~T in 2012) magnetic field~\cite{2022HuangNST33}. The solenoid is supported by an octagonal flux-return yoke with resistive plate counter muon identification modules interleaved with steel. The charged-particle momentum resolution at $1~{\rm GeV}/c$ is $0.5\%$, and the ${\rm d}E/{\rm d}x$ resolution is $6\%$ for electrons from Bhabha scattering. The EMC measures photon energies with a resolution of $2.5\%$ ($5\%$) at $1$~${\rm GeV}$ in the barrel (end-cap) region. The time resolution in the TOF barrel region is 68~ps, while that in the end-cap region is 110~ps. The end cap of the TOF system was upgraded in 2015 using the multi-gap resistive plate chamber technology, providing a time resolution of
60~ps~\cite{1720LGCRN1953}.

Simulated data samples produced with a {\sc geant4}-based~\cite{geant4} Monte Carlo (MC) package, which includes the geometric description of the BESIII detector and the detector response, are used to determine detection efficiencies and to estimate backgrounds. The simulation procedure models the beam energy spread and initial state radiation in the $e^+e^-$ annihilations with the generator {\sc kkmc}~\cite{ref:kkmc}.
The inclusive MC sample includes both the production of the $J/\psi$ resonance and the continuum processes incorporated in {\sc
kkmc}~\cite{ref:kkmc}. All particle decays are modelled with {\sc evtgen}~\cite{ref:evtgen} using branching fractions either taken from the Particle Data Group~\cite{2022PdgPTEP2022}, when available, or otherwise estimated with {\sc lundcharm}~\cite{ref:lundcharm}. Final state radiation from charged final state particles is incorporated using the {\sc photos} package~\cite{photos}. To determine the detection efficiency, the signal MC samples with $J/\psi \rightarrow \Xi^{0}\bar{\Xi}^{0}$, $\bar{\Xi}^{0} \rightarrow \bar{\Lambda}\pi^{0}$, and $\Xi^{0} \rightarrow K^{\pm} e^{\mp}$ are produced, where the $J/\psi$ and $\bar{\Xi}^{0} \rightarrow \bar{\Lambda}\pi^{0}$ decays are generated with the measured parameters in Refs.~\cite{2017ParaPLB770, 2022PdgPTEP2022}, and the signal decays of $\Xi^{0} \rightarrow K^{\pm} e^{\mp}$ are generated with a uniform phase space distribution.

\section{\boldmath{EVENT SELECTION AND DATA ANALYSIS}}
\label{evt_sel}

\subsection{\boldmath{Analysis method}}
\label{method}

The $\Xi^{0}\text{-}\bar{\Xi}^{0}$ hyperons are produced in pairs from $J/\psi$ decays without any additional fragmentation particles. This property provides an ideal environment for the double-tag (DT) method, which was first introduced by the MARK-III collaboration~\cite{1986DTPRL56}. In this approach, one $\bar{\Xi}^{0}$ hyperon is fully reconstructed via its dominant decay mode $\bar{\Xi}^{0} \rightarrow \bar{\Lambda}(\rightarrow \bar{p}\pi^{+})\pi^{0}(\rightarrow \gamma\gamma)$, and referred to as ``tagged $\bar{\Xi}^{0}$". Then the signal decays $\Xi^{0} \rightarrow K^{\pm} e^{\mp}$ are searched for in the recoiling side of the tagged $\bar{\Xi}^{0}$. The tagged $\bar{\Xi}^{0}$ candidates are referred to as ``single-tag" (ST) candidates, while the events in which the signal decays of interest and the tagged $\bar{\Xi}^{0}$ are simultaneously found are referred to as DT events. The absolute branching fraction of the signal decay $\mathcal{B}_{\rm sig}$ is calculated by
\begin{equation}
\label{eq:br_sig}
  \mathcal{B}_{\rm sig} = \frac{N_{\rm DT}^{\rm obs}}{N_{\rm ST}^{\rm obs}\cdot\epsilon_{\rm DT}/\epsilon_{\rm ST}},
\end{equation}
where $N_{\rm ST}^{\rm obs}$ ($N_{\rm DT}^{\rm obs}$) is the ST (DT) yield, and $\epsilon_{\rm ST}$ ($\epsilon_{\rm DT}$) is the ST (DT) detection efficiency.

\subsection{\boldmath{ST selection}}
\label{ST_sel}

Charged tracks detected in the MDC are required to be within a polar angle ($\theta$) range of $|\rm{cos\theta}|<0.93$, where $\theta$ is defined with respect to the $z$ axis, which is the symmetry axis of the MDC. Particle identification~(PID) for charged tracks combines measurements of the ${\rm d}E/{\rm d}x$ in the MDC and the flight time in the TOF. The PID confidence levels are calculated for the proton, pion, and kaon hypotheses. Tracks are identified as protons (pions) when the proton (pion) hypothesis has the highest confidence level among these three hypotheses.

The photon candidates are identified using their showers in the EMC. The deposited energy of each shower must be more than 25~${\rm MeV}$ in the barrel region ($|\!\cos\theta|<0.80$) and more than 50~${\rm MeV}$ in the end-cap region ($0.86<|\!\cos\theta|<0.92$). To exclude showers that originate from charged tracks, the angle subtended by the EMC shower and the position of the closest charged track at the EMC must be greater than 10 degrees as measured from the interaction point. To suppress electronic noise and showers unrelated to the event, the difference between the EMC time and the event start time is required to be within [0, 700] ns. The $\pi^{0}$ candidates are reconstructed with a pair of photons. Due to the poor resolution in the end-cap regions of the EMC, the $\pi^{0}$ candidates with two daughter photons found in the end caps are rejected. The invariant mass of the two photons is required to be within $(0.115, 0.150)$~${\rm GeV}/c^{2}$. A kinematic fit is performed by constraining the invariant mass of $\gamma\gamma$ to the known $\pi^{0}$ mass~\cite{2022PdgPTEP2022}.

To reconstruct $\bar{\Lambda}$ candidates, a vertex fit~\cite{2009XuCPC33} is applied to all $\bar{p}\pi^{+}$ combinations, and the one closest to the known $\bar{\Lambda}$ mass ($M_{\bar{\Lambda}}$)~\cite{2022PdgPTEP2022} is retained for further analysis. The invariant mass of $\bar{p}\pi^{+}$ is required to satisfy $|M_{\bar{p}\pi^{+}}-M_{\bar{\Lambda}}|<0.005$~${\rm GeV}/c^{2}$. The $\bar{\Xi}^{0}$ candidates are reconstructed with the $\bar{\Lambda}\pi^{0}$ combinations, and the one closest to the known $\bar{\Xi}^{0}$ mass ($M_{\bar{\Xi}^{0}}$)~\cite{2022PdgPTEP2022} is retained for further analysis. The invariant mass of $\bar{\Lambda}\pi^{0}$ is required to satisfy $|M_{\bar{\Lambda}\pi^{0}}-M_{\bar{\Xi}^{0}}|<0.02$~${\rm GeV}/c^{2}$, which corresponds to about three times the resolution ($\pm 3\sigma$) around $M_{\bar{\Xi}^{0}}$.

\begin{figure}[tb]
\centering
\includegraphics[width=0.45\textwidth]{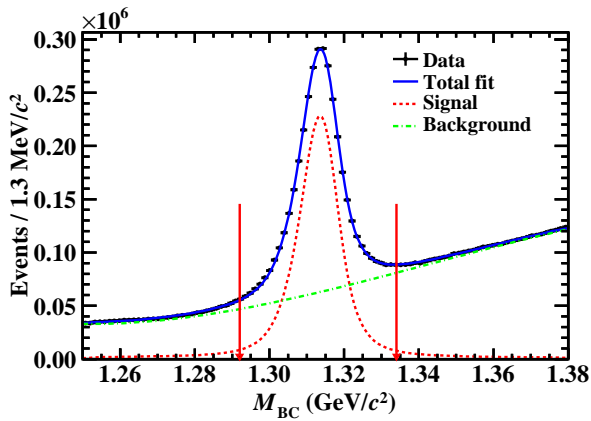}
\caption{Fit to the $M_{\rm BC}$ distribution of the accepted ST candidates, where the black points with error bars are data, and the red dashed and green dash-dotted lines are the signal and background shapes, respectively. The blue solid line shows the total fit and the red arrows indicate the ST signal region.}
\label{fig:ST_mxi0}
\end{figure}

The ST yield of $\bar{\Xi}^{0}$ hyperons is extracted from a binned maximum likelihood fit to the distribution of a beam-constrained mass defined as
\begin{equation}
\label{eq:mbc}
  M_{{\rm BC}} = \sqrt{E_{{\rm beam}}^{2}-|\vec{p}_{\bar{\Lambda}\pi^{0}}|^{2}},
\end{equation}
where $E_{\rm beam}$ is the beam energy, and $\vec{p}_{\bar{\Lambda}\pi^{0}}$ is the momentum of the selected $\bar{\Lambda}\pi^{0}$ combination in the center-of-mass system. In the fit, the signal shape is modeled by the MC-simulated shape convolved with a Gaussian function to describe the resolution difference between data and MC simulation. By analyzing the inclusive MC sample with a generic event type analysis tool, TopoAna~\cite{2021ZhouCPC258}, we find that there is no peaking background. Therefore, the background shape is described with a third-order Chebychev polynomial function. Figure~\ref{fig:ST_mxi0} shows the best fit result, and candidates in the ST signal region of $1.292<M_{\rm BC}<1.334~{\rm GeV}/c^{2}$, which corresponds to about $\pm 3\sigma$ around $M_{\bar{\Xi}^{0}}$, are kept for further analysis. From the fit, the ST yield is obtained to be $2,538,372\pm 2593$. The ST efficiency is determined to be $(16.26\pm0.02)\%$ by performing the same analysis strategy as that used in the data analysis to the inclusive MC sample. The uncertainties of these two numbers are statistical only.

\subsection{\boldmath{DT selection}}
\label{DT_sel}
The signal candidates for $\Xi^{0} \rightarrow K^{\pm} e^{\mp}$ are selected from the remaining tracks recoiling against the tagged $\bar{\Xi}^{0}$.
The PID confidence levels are calculated with the combined ${\rm d}E/{\rm d}x$, TOF and EMC information for the electron or positron (${\rm CL}_{e}$), kaon (${\rm CL}_{K}$) and pion (${\rm CL}_{\pi}$) hypotheses. The track is identified as an electron or positron if it satisfies ${\rm CL}_{e}>0.1\%$ and ${\rm CL}_{e}/({\rm CL}_{e}+{\rm CL}_{\pi}+{\rm CL}_{K})>0.8$. The track is identified as a kaon when the kaon hypothesis has the highest confidence level. The $\Xi^{0}$ is reconstructed with the $K^{\pm} e^{\mp}$ combinations, and the one with the invariant mass $M_{Ke}$ closest to the known $\Xi^{0}$ mass ($M_{\Xi^{0}}$)~\cite{2022PdgPTEP2022} is retained for further analysis. To further suppress backgrounds, the opening angle between the $\Xi^{0}$ and $\bar{\Xi}^{0}$ momenta is required to satisfy $\theta_{\Xi^0\bar{\Xi}^{0}}>178.5^{\circ}$, which is optimized using the Punzi figure of merit with a formula of $\epsilon/(0.8+\sqrt{{\rm B}})$~\cite{2003Punzi}, where $\epsilon$ denotes the signal efficiency, ``B" is the number of background events estimated from the inclusive MC sample, and the number ``0.8" is half of the sigma number corresponding to the desired confidence level (90\%).

\begin{figure}[tb]
\centering
\includegraphics[width=0.45\textwidth]{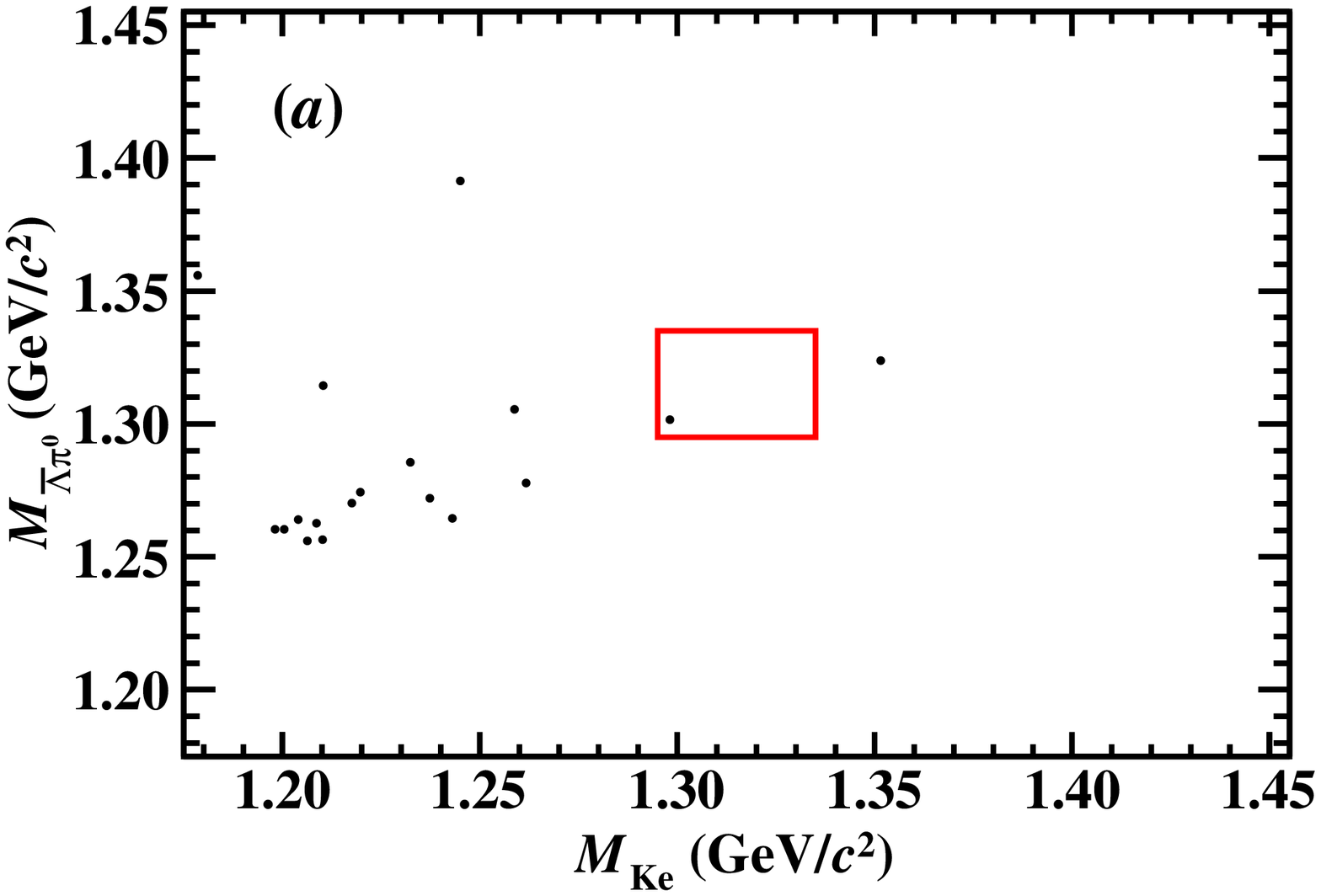}\\
\includegraphics[width=0.45\textwidth]{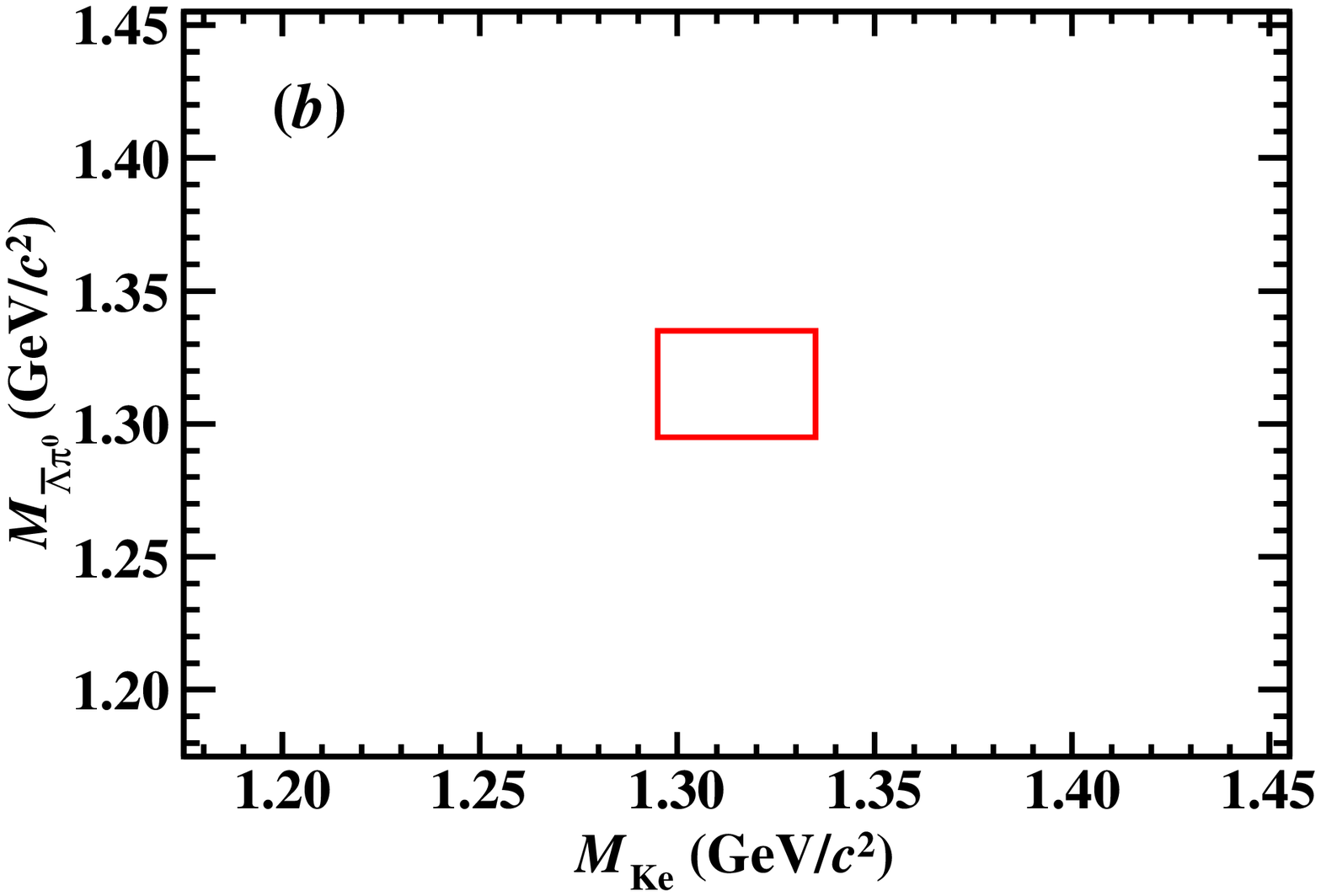}
\caption{Distributions of $M_{Ke}$ versus $M_{\bar{\Lambda}\pi^{0}}$ of the accepted candidates for (a) $\Xi^{0}\rightarrow K^{-}e^{+}$ and (b) $\Xi^{0}\rightarrow K^{+}e^{-}$ in data, respectively. The red box indicates the signal region.}
\label{fig:scatter}
\end{figure}

The DT yield is determined with the distribution of $M_{Ke}$ versus $M_{\bar{\Lambda}\pi^{0}}$ of the accepted candidates in data, as shown in Fig.~\ref{fig:scatter}, where $M_{Ke}$ ($M_{\bar{\Lambda}\pi^{0}}$) is the invariant mass of the $K^{\pm} e^{\mp}$ ($\bar{\Lambda}\pi^{0}$) combination. The signal region is defined as $|M_{\bar{\Lambda}\pi^{0}}-M_{\bar{\Xi}^{0}}|<0.02~{\rm GeV}/c^{2}$ and $|M_{Ke}-M_{\Xi^{0}}|<0.02~{\rm GeV}/c^{2}$, which correspond to about $\pm 3\sigma$ around $M_{\Xi^{0}}$. One event is observed in the signal region of $\Xi^{0}\rightarrow K^{-}e^{+}$, while no event is observed in the signal region of $\Xi^{0}\rightarrow K^{+}e^{-}$. By analyzing the inclusive MC sample, we find that there is no background event left in these two signal regions. Considering the very limited statistics, we set the upper limit on the DT yield as described in Sec.~\ref{result}. The DT efficiency of $\Xi^{0}\rightarrow K^{-}e^{+}$ obtained from the signal MC sample is $(6.64\pm0.01)\%$, and the uncertainty here is only statistical, while the $\Xi^{0}\rightarrow K^{+}e^{-}$ has the same DT efficiency as $\Xi^{0}\rightarrow K^{-}e^{+}$.

\section{\boldmath SYSTEMATIC UNCERTAINTIES}
\label{sys_err}

The systematic uncertainties mainly originate from the tracking efficiency, PID efficiency, ST fit, tag bias, and $\theta_{\Xi^0\bar{\Xi}^{0}}$ requirement, as summarized in Table~\ref{tab:sum_sys}. Most of the systematic uncertainties on the ST side cancel due to the DT method described in Sec.~\ref{method}.

The uncertainties due to the tracking efficiencies are determined to be 0.6\% for the kaon by studying the control sample of $J/\psi\rightarrow K_{\rm S}^{0}K^{\pm}\pi^{\mp}$, and 0.1\% for the electron or positron by studying the radiative Bhabha events $e^{+}e^{-}\rightarrow \gamma e^{+}e^{-}$ at $\sqrt{s}=3.097$ ${\rm GeV}$. The uncertainties arising from the PID efficiencies for kaon (0.4\%) and electron or positron (1.1\%) are assigned with the same control samples as used in the studies of tracking efficiencies. The uncertainty due to the ST fit is determined to be 2.8\% by varying the signal shape (2.6\%) to the one without convolving with the Gaussian function, changing the background shape (0.7\%) from the third-order Chebychev function to the second-order one, and altering the fit range (0.7\%) from (1.25, 1.38) ${\rm GeV}/c^{2}$ to (1.23, 1.40) ${\rm GeV}/c^{2}$. The uncertainty due to the tag bias arising from the difference of ST efficiencies in the inclusive and signal MC samples caused by different multiplicities is assigned to be 0.2\% by following the method described in Ref.~\cite{2021TagbiasPRD104}. To obtain the uncertainty arising from the $\theta_{\Xi^0\bar{\Xi}^{0}}$ requirement, a selection efficiency is defined by counting the number of events with and without the $\theta_{\Xi^0\bar{\Xi}^{0}}$ requirement, and the difference of selection efficiencies between data and MC simulation, 0.8\%, is assigned as the corresponding systematic uncertainty. The total systematic uncertainty is estimated to be 3.4\% by adding all these uncertainties in quadrature.

\begin{table}[htp]
\center
\caption{Relative systematic uncertainties for the branching fraction measurement.}
\label{tab:sum_sys}
\begin{tabular}{lc}
\hline\hline
Source                                                  & Uncertainty~(\%) \\ \hline
Tracking                                                & 0.7        \\
PID                                                     & 1.5        \\
ST fit                                                  & 2.8        \\
Tag bias                                                & 0.2        \\
$\theta_{\Xi^0\bar{\Xi}^{0}}$                           & 0.8        \\ \hline
Total                                                   & 3.4        \\ \hline \hline
\end{tabular}
\vspace{-0.2cm}
\end{table}

\section{\boldmath{RESULT}}
\label{result}

Since no obvious signal event is observed in the signal region, the upper limits on the DT yields of signal decays are set with a frequentist method of profile likelihood as a general treatment of nuisance parameters~\cite{2005TRolkeNIMA551}. The numbers of the signal events and background events are assumed to follow the Poisson distribution, the detection efficiency is assumed to follow a Gaussian distribution, and the systematic uncertainty is considered as the standard deviation of the efficiency. This method uses the numbers of observed events in the signal and background regions, ratio of the signal-region size to the background-region size, efficiency, systematic uncertainty and confidence level as input parameters to calculate the upper limit on the DT yield. Considering the choice of the background region can affect the upper limit on the DT yield, we sample the background region randomly with a Gaussian distribution in the range of $(1.175, 1.455)~{\rm GeV}/c^{2}$ for $M_{Ke}$ and $ M_{\bar{\Lambda}\pi^{0}}$ that excludes the signal region, and conservatively take the one giving the maximum upper limit on the DT yield as nominal background region. The upper limit on the DT yield of $\Xi^{0}\rightarrow K^{-}e^{+}$ ($\Xi^{0}\rightarrow K^{+}e^{-}$) is determined to be 3.7 (2.0) at the 90\% confidence level. Based on Eq.~(\ref{eq:br_sig}), the upper limits on the branching fractions at the 90\% confidence level are determined to be
\begin{equation}
\begin{split}
\label{eq:br_UL}
\mathcal{B}(\Xi^{0}\rightarrow K^{-}e^{+})<3.6\times10^{-6},\\
\mathcal{B}(\Xi^{0}\rightarrow K^{+}e^{-})<1.9\times10^{-6}.
\end{split}
\end{equation}

\section{\boldmath SUMMARY}
\label{summary}

In summary, based on $(1.0087\pm0.0044)\times10^{10}$ $J/\psi$ events collected at $\sqrt{s} = 3.097$ ${\rm GeV}$ with the BESIII detector, we present the first search for the baryon and lepton number violating decays of $\Xi^{0} \rightarrow K^{-} e^{+}$ with $\Delta(B-L)=0$ and $\Xi^{0} \rightarrow K^{+} e^{-}$ with $|\Delta(B-L)|=2$. No obvious signal is observed, and the upper limits on the branching fractions of these two decays are set to be $\mathcal B(\Xi^{0} \rightarrow K^{-} e^{+})<3.6\times10^{-6}$ and $\mathcal B(\Xi^{0} \rightarrow K^{+} e^{-})<1.9\times10^{-6}$ at the 90\% confidence level. These results are among the best constraints on the BNV interactions from hyperon decays. They offer a direct probe of the BNV processes involving strange quarks in the initial states.

\section*{\boldmath ACKNOWLEDGMENTS}

The BESIII Collaboration thanks the staff of BEPCII and the IHEP computing center for their strong support. This work is supported in part by National Key R\&D Program of China under Contracts Nos. 2020YFA0406400, 2020YFA0406300; National Natural Science Foundation of China (NSFC) under Contracts Nos. 11635010, 11735014, 11835012, 11935015, 11935016, 11935018, 11961141012, 12022510, 12025502, 12035009, 12035013, 12061131003, 12192260, 12192261, 12192262, 12192263, 12192264, 12192265; the Chinese Academy of Sciences (CAS) Large-Scale Scientific Facility Program; the CAS Center for Excellence in Particle Physics (CCEPP); Joint Large-Scale Scientific Facility Funds of the NSFC and CAS under Contract No. U1832207; CAS Key Research Program of Frontier Sciences under Contracts Nos. QYZDJ-SSW-SLH003, QYZDJ-SSW-SLH040; 100 Talents Program of CAS; The Institute of Nuclear and Particle Physics (INPAC) and Shanghai Key Laboratory for Particle Physics and Cosmology; ERC under Contract No. 758462; European Union's Horizon 2020 research and innovation programme under Marie Sklodowska-Curie grant agreement under Contract No. 894790; German Research Foundation DFG under Contracts Nos. 443159800, 455635585, Collaborative Research Center CRC 1044, FOR5327, GRK 2149; Istituto Nazionale di Fisica Nucleare, Italy; Ministry of Development of Turkey under Contract No. DPT2006K-120470; National Research Foundation of Korea under Contract No. NRF-2022R1A2C1092335; National Science and Technology fund of Mongolia; National Science Research and Innovation Fund (NSRF) via the Program Management Unit for Human Resources \& Institutional Development, Research and Innovation of Thailand under Contract No. B16F640076; Polish National Science Centre under Contract No. 2019/35/O/ST2/02907; The Royal Society, UK under Contract No. DH160214; The Swedish Research Council; U. S. Department of Energy under Contract No. DE-FG02-05ER41374. This paper is also supported by the Youth Science Fund of Henan Normal University under Contract No. 2021QK12, the NSFC under Contract No. 11947038, and the High Performance Computing Center of Henan Normal University.


\begin{thebibliography}{99}
\bibitem{1997AdamsRMP69}
  F. C. Adams and G. Laughlin, \href{https://doi.org/10.1103/RevModPhys.69.337}{Rev. Mod. Phys. {\bf 69}, 337 (1997)}.

\bibitem{1967SakharovPZETF5}
  A. D. Sakharov, \href{http://jetpletters.ru/ps/1643/article_25089.shtml}{JETP Lett. {\bf 5}, 32 (1967); {\bf 5}, 109 (E) (1967)}.

\bibitem{1979WeinbergPRL43}
  S. Weinberg, \href{https://doi.org/10.1103/PhysRevLett.43.1566}{Phys. Rev. Lett. {\bf 43}, 1566 (1979)}; F. Wilczek and A. Zee, \href{https://doi.org/10.1103/PhysRevLett.43.1571}{Phys. Rev. Lett. {\bf 43}, 1571 (1979)}; L. F. Abbott and M. B. Wise, \href{https://doi.org/10.1103/PhysRevD.22.2208}{Phys. Rev. D {\bf 22}, 2208 (1980)}.

\bibitem{1974GeorgiPRL32}
  H. Georgi and S. L. Glashow, \href{https://doi.org/10.1103/PhysRevLett.32.438}{Phys. Rev. Lett. {\bf 32}, 438 (1974)}.

\bibitem{2009SKPRL102}
  H. Nishino {\it et al.} (Super-Kamiokande Collaboration), \href{https://doi.org/10.1103/PhysRevLett.102.141801}{Phys. Rev. Lett. {\bf 102}, 141801 (2009)}.

\bibitem{1989IPLB231}
  I. Antoniadis, John Ellis, J. S. Hagelin, and D. V. Nanopoulos, \href{https://doi.org/10.1016/0370-2693(89)90115-9}{Phys. Lett. B {\bf 231}, 65 (1989)}.

\bibitem{1980WeinbergPRD22}
  S. Weinberg, \href{https://doi.org/10.1103/PhysRevD.22.1694}{Phys. Rev. D {\bf 22}, 1694 (1980)}; H.A. Weldon and A. Zee, \href{https://doi.org/10.1016/0550-3213(80)90218-7}{Nucl. Phys. B {\bf 173}, 269 (1980)}.

\bibitem{2012BabuPRL109}
  K. S. Babu and R. N. Mohapatra, \href{https://doi.org/10.1103/PhysRevLett.109.091803}{Phys. Rev. Lett. {\bf 109}, 091803 (2012)}.

\bibitem{2020DPRD101}
  M. Ablikim {\it et al.} (BESIII Collaboration), \href{https://doi.org/10.1103/PhysRevD.101.031102}{Phys. Rev. D {\bf 101}, 031102(R) (2020)};
  \href{https://doi.org/10.1103/PhysRevD.106.112009}{Phys. Rev. D {\bf 106}, 112009 (2022)}; \href{https://doi.org/10.1103/PhysRevD.105.032006}{Phys. Rev. D {\bf 105}, 032006 (2022)}.

\bibitem{2019JpsiPRD99}
  M. Ablikim {\it et al.} (BESIII Collaboration), \href{https://doi.org/10.1103/PhysRevD.99.072006}{Phys. Rev. D {\bf 99}, 072006 (2019)}.

\bibitem{2020tauPRD102}
  D. Sahoo {\it et al.} (Belle Collaboration), \href{https://doi.org/10.1103/PhysRevD.102.111101}{Phys. Rev. D {\bf 102}, 111101(R) (2020)}.

\bibitem{2011BPRD83}
  P. del Amo Sanchez {\it et al.} (BaBar Collaboration), \href{https://doi.org/10.1103/PhysRevD.83.091101}{Phys. Rev. D {\bf 83}, 091101(R) (2011)}.

\bibitem{2014TopPLB731}
  S. Chatrchyan {\it et al.} (CMS Collaboration), \href{https://doi.org/10.1016/j.physletb.2014.02.033}{Phys. Lett. B {\bf 731}, 173 (2014)}.

\bibitem{2020SKPRD102}
  A. Takenaka {\it et al.} (Super-Kamiokande Collaboration), \href{https://doi.org/10.1103/PhysRevD.102.112011}{Phys. Rev. D {\bf 102}, 112011 (2020)}.

\bibitem{2005HouPRD72}
  W. S. Hou, M. Nagashima, and A. Soddu, \href{https://doi.org/10.1103/PhysRevD.72.095001}{Phys. Rev. D {\bf 72}, 095001 (2005)}.

\bibitem{2015LambdaPRD92}
  M. E. McCracken {\it et al.} (CLAS Collaboration), \href{https://doi.org/10.1103/PhysRevD.92.072002}{Phys. Rev. D {\bf 92}, 072002 (2015)}.

\bibitem{2017LiFP12}
  H. B. Li, \href{https://doi.org/10.1007/s11467-017-0691-9}{Front. Phys. \textbf{12}, 121301 (2017)}; \href{https://doi.org/10.1007/s11467-019-0910-7}{\textbf{14}, 64001(E) (2019)}.

\bibitem{2022YangCPC46}
   M. Ablikim {\it et al.} (BESIII Collaboration), \href{https://dx.doi.org/10.1088/1674-1137/ac5c2e}{Chin. Phys. C {\bf 46}, 074001 (2022)}.

\bibitem{2009AblikimNIMA614}
  M. Ablikim {\it et al.} (BESIII Collaboration),
  \href{https://doi.org/10.1016/j.nima.2009.12.050}{Nucl. Instrum. Methods Phys. Res., Sect. A {\bf 614}, 345 (2010)}.

\bibitem{2016YuIPACTUYA01}
   C. H. Yu {\it et al.}, \href{https://doi.org/10.18429/JACoW-IPAC2016-TUYA01}{in {\it Proceedings of IPAC2016, Busan, Korea} (2016)}.

\bibitem{2019AblikimCPC44}
  M. Ablikim {\it et al.} (BESIII Collaboration),
  \href{https://doi.org/10.1088/1674-1137/44/4/040001}{Chin. Phys. C {\bf 44}, 040001 (2020)}.

\bibitem{2022HuangNST33}
  K.~X.~Huang {\it et al.},
  \href{https://doi.org/10.1007/s41365-022-01133-8}{Nucl. Sci. Tech. {\bf 33}, 142 (2022)}.

\bibitem{1720LGCRN1953}
  X. Li {\it et al.}, \href{https://doi.org/10.1007/s41605-017-0014-2}{Radiat. Detect. Technol. Methods {\bf 1}, 13 (2017)};
  Y. X. Guo {\it et al.}, \href{https://doi.org/10.1007/s41605-017-0012-4}{Radiat. Detect. Technol. Methods {\bf 1}, 15 (2017)};
  P. Cao {\it et al.}, \href{https://doi.org/10.1016/j.nima.2019.163053}{Nucl. Instrum. Methods Phys. Res., Sect. A {\bf 953}, 163053 (2020)}.

\bibitem{geant4}
  S.~Agostinelli {\it et al.} (GEANT4 Collaboration),
  \href{https://doi.org/10.1016/S0168-9002(03)01368-8}{Nucl. Instrum. Methods Phys. Res., Sect. A {\bf 506}, 250 (2003)}.

\bibitem{ref:kkmc}
  S.~Jadach, B.~F.~L.~Ward, and Z.~Was,
  \href{https://doi.org/10.1103/PhysRevD.63.113009}{Phys.\ Rev.\ D {\bf 63}, 113009 (2001)};
  \href{https://doi.org/10.1016/S0010-4655(00)00048-5}{Comput.\ Phys.\ Commun.\  {\bf 130}, 260 (2000)}.

\bibitem{ref:evtgen}
  D.~J.~Lange,
  \href{https://doi.org/10.1016/S0168-9002(01)00089-4}{Nucl. Instrum. Methods Phys. Res., Sect. A {\bf 462}, 152 (2001)};
  R.~G.~Ping,
  \href{https://doi.org/10.1088/1674-1137/32/8/001}{Chin. Phys. C {\bf 32}, 599 (2008)}.

\bibitem{2022PdgPTEP2022}
  R.~L.~Workman {\it et al.} (Particle Data Group),
  \href{https://doi.org/10.1093/ptep/ptac097}{Prog. Theor. Exp. Phys. \textbf{2022}, 083C01 (2022)}.

\bibitem{ref:lundcharm}
  J.~C.~Chen, G.~S.~Huang, X.~R.~Qi, D.~H.~Zhang, and Y.~S.~Zhu,
  \href{https://doi.org/10.1103/PhysRevD.62.034003}{Phys.\ Rev.\ D {\bf 62}, 034003 (2000)};
  R.~L.~Yang, R.~G.~Ping, and H.~Chen,
  \href{https://doi.org/10.1088/0256-307x/31/6/061301}{Chin.\ Phys.\ Lett.\  {\bf 31}, 061301 (2014)}.

\bibitem{photos}
  E.~Richter-Was,
  \href{https://doi.org/10.1016/0370-2693(93)90062-M}{Phys.\ Lett.\ B {\bf 303}, 163 (1993)}.

\bibitem{2017ParaPLB770}
  M. Ablikim {\it et al.} (BESIII Collaboration),
  \href{https://doi.org/10.1016/j.physletb.2017.04.048}{Phys. Lett. B {\bf 770}, 217 (2017)}.

\bibitem{1986DTPRL56}
  R. M. Baltrusaitis {\it et al.} (MARK-III Collaboration),
  \href{https://doi.org/10.1103/PhysRevLett.56.2140}{Phys. Rev. Lett. \textbf{56}, 2140 (1986)}.

\bibitem{2009XuCPC33}
  M. Xu {\it et al.}, \href{https://doi.org/10.1088/1674-1137/33/6/005}{Chin. Phys. C \textbf{33}, 428 (2009)}.

\bibitem{2021ZhouCPC258}
  X.~Y.~Zhou, S.~X.~Du, G.~Li, and C.~P.~Shen,
  \href{https://doi.org/10.1016/j.cpc.2020.107540}{Comput. Phys. Commun. \textbf{258}, 107540 (2021)}.

\bibitem{2003Punzi}
  G. Punzi, \href{https://arxiv.org/abs/physics/0308063}{eConf \textbf{C030908}, MODT002 (2003)}.

\bibitem{2021TagbiasPRD104}
  M. Ablikim {\it et al.} (BESIII Collaboration),
  \href{https://doi.org/10.1103/PhysRevD.104.032001}{Phys. Rev. D \textbf{104}, 032001 (2021)}.

\bibitem{2005TRolkeNIMA551}
  W. A. Rolke, A. M. Lopez, and J. Conrad, \href{https://doi.org/10.1016/j.nima.2005.05.068}{Nucl. Instrum. Methods Phys. Res., Sect. A {\bf 551}, 493 (2005)};
  J. Conrad and J. Lundberg, \href{https://root.cern.ch/root/html/TRolke.html}{https://root.cern.ch/ root/html/TRolke.html}.

\end{thebibliography}
\end{document}